\documentclass[english,a4paper]{article}
\usepackage{epsfig,graphicx,times,amsmath,amsthm, amssymb, multicol}

\usepackage{epstopdf}

\begin{document}

\title {Determination of Rate Constant of Chemical Reactions by   Simple Numerical Nonlinear Analysis }
\author{Christopher G. Jesudason \\
{\normalsize Dept. of Chemistry and Center for Theoretical and Computational Physics,} \\
 {\normalsize University of Malaya, 50603 Kuala Lumpur}\\
 {\normalsize Malaysia}\\
 {\normalsize {\em E-mails:}jesu\symbol{64}um.edu.my, chrysostomg\symbol{64}gmail.com} }
 \vspace{.5cm}
 \maketitle
\begin{abstract}
 For some centuries, first order chemical rate constants  were determined mainly  by a linear logarithmic plot of  reagent concentration terms  against time where the initial concentration was required, which is experimentally often a  challenging task to derive accurate estimates. By definition, the rate constant was deemed to be invariant and the kinetic equations were developed with this assumption.   A reason for these developments was the ease in which linear graphs could be plotted.  Here, different methods are discussed that does not require  exact knowledge of initial concentrations  and which require elementary  nonlinear analysis and  the ensuing results are  compared with those derived from the standard methodology 	from an actual chemical reaction, with its experimental  determination of the initial concentration with a degree of uncertain. We verify experimentally  our previous theoretical conclusion  based on simulation data [ J. Math . Chem {\bf 43} (2008) 976--1023] that the so called  rate constant is never  constant even for elementary reactions and that all the rate laws and experimental determinations to date are actually averaged quantities over the reaction pathway. We conclude that nonlinear methods  in conjunction with experiments  could in the future play  a crucial role in extracting  information of various kinetic parameters.  
\end{abstract}

\section{1. INTRODUCTION AND METHODS} 
As  alluded in the abstract, most kinetic determinations use logarithmic plots with known initial concentrations, although there have been attempts \cite[and refs. therein]{moore1,gug1}. (There are possible ambiguities  in \cite{moore1} concerning  choice of variables that will not be discussed.) However  all these publications hitherto assume constancy of the rate constant $k$ and  do not focus on nonlinear analysis (NLA), as will be attempted here in preliminary form. We analyze kinetic data of the tert butyl chloride  hydrolysis reaction in ethanol solvent (80\%v/v) derived  from the Year III teaching laboratory of this University (UM); 0.3mL of the reactant was dissolved in 50mL of ethanol initially. The reaction was conducted at $30^oC$ and  monitored over time (minutes) by measuring conductivity ($\mu\text{S cm}^{-1}$) due to the release of $\text{H}^+$ and $\text{Cl}^-$ ions as shown below (\ref{eq:1}),
 \begin{equation} \label{eq:1}
\mbox{C$_4$H$_9$Cl}+\mbox{H$_2$O}\,\, \stackrel{k}{\longrightarrow}\,\,  \mbox{C$_4$H$_9$OH} + \mbox{H}^{+} + \mbox{Cl}^{-}
\end{equation}
 and $\lambda_\infty=2050 \mu\text{S cm}^{-1}$ was determined by heating the reaction vessel at the end of the monitoring to $60^o$C until there was no apparent change in the conductivity when equilibrated back at $30^o$C. ``Units'' in the figures and text refers to $\mu\text{S cm}^{-1}$. It would be inferred here that either because of evaporation or the temperatures not equilibrating after heating, the measured $\lambda_\infty$ is larger than the actual one. Linear proportionality is assumed in $\lambda$  and the extent of reaction $x$, where  the first order  law (c being  the instantaneous concentration and $a$ the initial concentration)   is $\frac{dc}{dt}=-kc=-k(a-x)$; with $\lambda_{\infty}=\alpha a,\lambda_{t}=\alpha x$ and $\lambda(0)=\lambda_{0}=\alpha x_0$, integration yields for assumed constant $k$
\begin{equation} \label{eq:2}
\ln\frac{(\lambda_\infty- \lambda_0)}{(\lambda_\infty-\lambda(t))}=kt
\end{equation}
Eqn.(\ref{eq:2}) determines $k$ if $\lambda_{0}$ and $\lambda_{\infty}$  are known.
\begin{figure}[htbp]
\begin{center}
\includegraphics[ width=7cm]{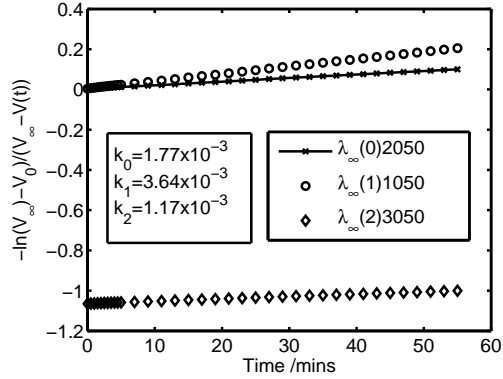} 
\end{center}
\caption{Integrated equation(\ref{eq:2}) plot with $\lambda_{\infty}$ from experiment (0) and from two different arbitrary values (1,2) for $\lambda_{\infty}$}
\label{fig:1} 
\end{figure}
The plot of (\ref{eq:2}) was made for  the same experimental values with different $\lambda_\infty$'s, both higher and lower than the experimental value. We find that the rate constant for the NLA  was higher, leading to a lower value of $\lambda_\infty$ which is consonant with evaporation of solvent or the non-equilibration of temperature prior to measurement  to determine $\lambda_{\infty}$.Except for the last subsection, we shall do a NLA based on constant $k$ assumption.
\subsection{1.1 Method 1} 
\begin{figure}[htbp]
\begin{center}
\includegraphics[width=7cm]{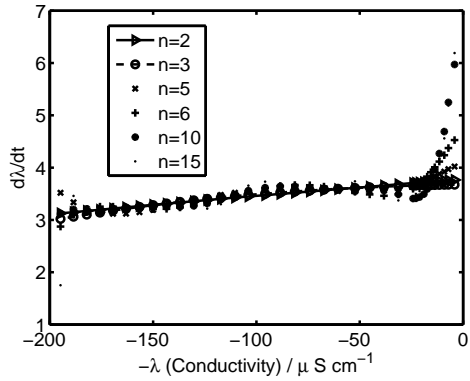} 
\end{center}
\caption{Method 1 graph showing  linearity  lower order polynomial fits}
\label{fig:2} 
\end{figure}

Under linearity argument and constant $k$, the rate equation $\frac{dc}{dt}=-kc=-k(a-x)$ reduces to 
\begin{equation} \label{eq:2b}
\frac{\lambda(t)}{dt}= -k\lambda(t)+\lambda_{\infty}.k
\end{equation}
Hence a  plot of $\frac{\lambda(t)}{dt}$ vs $t$ would be linear. We find this to be the case for polynomial order $npoly\leq 3$ as in Fig.(\ref{fig:2}) below for all data values; higher polynomial orders can be used in selected data points  of the  curve below, especially in the central region. Thus criteria must be set up to determine the appropriate regime of datapoints in the NLA.

\subsection{1.2 Method 2}
\begin{figure}[htbp]
\begin{center}
\includegraphics[width=7cm]{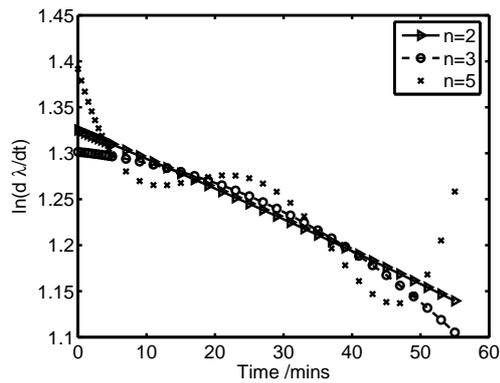} 
\end{center}
\caption{Method 2 where smooth curves are obtained for  at least $npoly<4$ }
\label{fig:3} 
\end{figure}

Let $\alpha'=\lambda_\infty-\lambda_0$, then  $\ln \alpha' -\ln(\lambda_\infty-\lambda)=kt$, then noting this and differentiating  yields 
\begin{equation} \label{eq:3}
\underbrace{\ln\left(\frac{d\lambda}{dt}\right)}_{Y}=\underbrace{-kt}_{Mt}+\underbrace{\ln[k(\lambda_\infty- \lambda_0)]}_{C}
\end{equation}
A typical plot that can extract $k$ as a linear plot of $\ln(d\lambda/dt)$ vs $t$ is given in Fig.(\ref{fig:3}).  Linearity  is observed  for $npoly=2$ and smooth curves without oscillations for at least $npoly\leq3$.  
\subsection{1.3 Other methods and considerations}
A variant  method similar to the Guggenheim method \cite{gug1} of elimination is given below but where gradients to the conductivity curve is required, and where the average over all pairs is required.
\begin{equation} \label{eq:4}
\left\langle k\right\rangle=\frac{-2}{N(N-1)}\sum_{i}^{N}\sum_{j>i}^N\ln\left(\lambda'(t_i)/\lambda'(t_j)\right)/
(t_i-t_j)
\end{equation}
It was discovered that the normal least squares polynomial method using Gaussian elimination \cite[Sec.6.2.4,p.318 ]{yak1}to derive the coefficients of the polynomial was highly unstable for $npoly>4$ and so for this  work, we used a variant of the Orthogonal method modified for determination of differentials. The normal method defines the nth order polynomial $p_n(t)$ which is then expressed as a sum of square terms over the domain of measurement to yield $Q$ in eqns(\ref{eq:5}).
\begin{equation} \label{eq:5}
\begin{array}{rcl}
p_n(t) &=& \sum_{j=0}^{n}h_it^j\\
Q(f,p_n) &=&  \sum_{i=1}^N\left[f_i-p_n(t_i)\right]^2
	\end{array}
\end{equation}
The $Q$ function is minimized over the polynomial coefficient space. In the Orthogonal method adopted here, we  express  our polynomial expression $p_m(t)$ linearly in  coefficients $a_j$  of $\varphi_j$  functions that are orthogonal with respect to an {\it inner} product definition. For  arbitrary functions $f,g$, the inner product $(f,g)$  is defined below, together with properties of the $\varphi_j$ orthogonal polynomials.  

\begin{equation} \label{eq:6}
\begin{array}{rcl}
(f,g) &=& \sum_{k=1}^{N} f(t_k).g(t_k)\\
 (\varphi_i,\varphi_j)=0  &(i\neq j);&  \,\, \mbox{and} \,\, (\varphi_i,\varphi_i)\neq 0.
	\end{array}
\end{equation}

\begin{equation} \label{eq:7}
\begin{array}{rll}
\varphi_i(t)&=&(t-b_i)\varphi_{i-1}(t)-c_i\varphi_{i-2}(t)\,(i\geq1)\\
 \varphi_0(t)&=&1,\mbox{and}\,\, \varphi_j=0  \,\,\, j<1,\\
 b_i &=& (t\varphi_{i-1},\varphi_{i-1})/( \varphi_{i-1}, \varphi_{i-1}) \,\, (i\geq1) \\
 c_i &=&(t\varphi_{i-1},\varphi_{i-2})/( \varphi_{i-2}, \varphi_{i-2}) \,\, (i\geq2), c_i=0. 
	\end{array}
\end{equation}
We define the $m^{th}$ order polynomial  and associated $a_j$ coefficients as:

\begin{equation} \label{eq:8}
\begin{array}{rll}
p_m(t)&=& \sum_{j=0}^m a_j\varphi_{j}(t)\\[0.5cm]
a_j &=&(f, \varphi_{j})/(\varphi_{j}, \varphi_{j}), (j=0,1,\ldots m) \\
 	\end{array}
\end{equation}

\begin{figure}[htbp]
\begin{center}
\includegraphics[width=7cm]{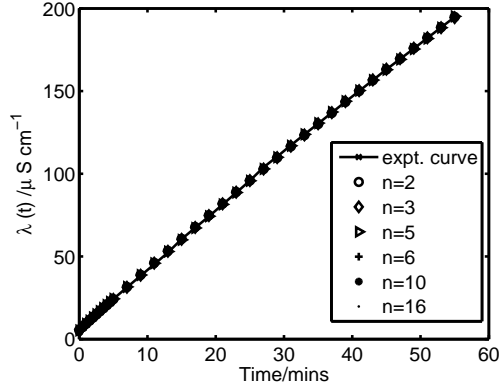} 
\end{center}
\caption{Plot using orthogonal polynomials for various orders $n$}
\label{fig:4} 
\end{figure}

The recursive definitions for the first and second derivatives are given respectively as:
\begin{equation} \label{eq:9}
\begin{array}{rll}
\varphi^{\prime}_i(t)&=&\varphi^{\prime}_{i-1}(t) (t-b_i)+\varphi_{i-1}(t)-c_i\varphi^{\prime}_{i-2}(t)\,(i\geq1)\\
\varphi^{\prime\prime}_i(t)&=&\varphi^{\prime\prime}_{i-1}(t) (t-b_i)+2\varphi_{i-1}^{\prime}(t)-c_i\varphi^{\prime\prime}_{i-2}(t)\,(i\geq2)
\end{array}
\end{equation}
Here the codes were developed in C/C++ which provides for recursive functions which we exploited for the evaluation of all the terms. The experimental data were fitted to an $m^{th}$ order expression  $\lambda_m(t)$ defined below 
\begin{equation} \label{eq:10}
\lambda_m(t) = \sum_{j=0}^{n}h_it^j\\
\end{equation}
Figure(\ref{fig:4})are plots for the different polynomial orders n. The orthogonal polynomial method is stable  and the mean square error decreases with higher polynomial order (for the 36 data points) monotonically, but the differentials are not so stable, as shown in the previous figures. 

Differentiating (\ref{eq:3}) for constant $k$ leads to (\ref{eq:11}) expressed in two ways
\begin{equation} \label{eq:11}
\frac{d^2 \lambda}{dt^2}=-k\left(\frac{d\lambda}{dt}\right)\,(a) \,\mbox{or}\,
k=-\frac{d^2 \lambda}{dt^2}/\left(\frac{d\lambda}{dt}\right)\,(b)
\end{equation}

 Eq.(\ref{eq:11}(b)) suggests another way of computing $k$ for ``well-behaved'' values of the differentials, meaning regions where $k$ would appear to be a reasonable constant. The (a) form suggests an exponential solution. Define $\frac{d\lambda}{dt}\equiv dl$ and $\frac{d^2\lambda}{dt^2}\equiv d2l$. Then $dl(t)=A\exp(-kt)$ and $dl(0)=A=h_2$ from (\ref{eq:10}).Furthermore,  as $t\rightarrow 0$, $k=\left(-2h_2/h_1\right)$ and a global definition of the rate constant becomes possible based on the total system $\lambda(t)$ curve.
 
With a slight change of notation, we now define $dl$ and $d2l$ as referring to the continuous functions $dl(t)=A\exp(-kt)$ and $d2l(t)=-kA\exp(-kt)$ and we consider $(d\lambda/dt)$ and $d^2\lambda/dt^2 $ to belong to the values (\ref{eq:10}) derived from ls fitting where $(d\lambda/dt)=\lambda_m^{\prime}$, \,\, $(d^2\lambda/dt^2)=\lambda_m^{\prime\prime}$ which are the  experimental values for a curve fit of order $m$. From the experimentally derived gradients and differentials, we can define two non-negative functions $R_a(k)$ and $R_b(k)$ as below:
 \begin{equation} \label{eq:12}
 \begin{array}{rll}
 R_a(k)&=&\sum^N_{i=1}\left(\frac{d^2\lambda(t_i)}{dt^2}+kdl(t_i)\right)^2\\[.3cm]
 R_b(k)&=&\sum^N_{i=1}\left(\frac{d\lambda(t_i)}{dt}-dl(t_i)\right)^2\\
 &\mbox{where}&  \\
 f_a(k)=R^{\prime}_a(k) &\mbox{and}& f_b(k)=R^{\prime}_b(k)
 \end{array}
\end{equation}
 and a minimum exists at $f_a(k)=f_b(k)=0.$ We solve the equations $f_a$ , $f_b$  for their roots in k using the Newton-Raphson method and compute the rate constant $k$. The error threshold in the Newton-Raphson method was set at $\epsilon=1.0\times10^{-7}$ We provide a series of data of the form $\left[n,A,k_a,k_b,\lambda_{a,\infty},\lambda_{b,\infty}\right]$ where $n$ refers to the polynomial degree, $A$ the initial value constant as above, $k_a$  and $k_b$ is the rate constant for  function $f_a$ and $f_b$  (solved when the functions are zero respectively ) and 
 likewise  for $\lambda_{a,\infty}$ and $\lambda_{b,\infty}$. The $e$ symbol refers to base $10$ (decimal) exponents. The $\lambda_{\infty}$ values are averaged over  all the (36) data points from  the equation 
 \begin{equation} \label{eq:12a}
 \lambda_{\infty}=\frac{d\lambda(t)}{dt}\frac{1}{k}+ \lambda(t)
\end{equation}
The results are as follows:\\
$\left[2, 3.7634e0, 3.2876e-3, 3.2967e-3, 1.1506e3, 1.1477e3 \right]$,\\
$\left[3, 3.6745e0, 2.7537e-3, 2.7849e-3, 1.34756e3,1.3334e3\right]$,\\
$\left[4, 3.6380e0, 2.0973e-3, 2.4716e-3, 1.7408e3, 1.4900e3\right]$,\\
$\left[5, 4.0210e0, 9.7622e-3, 4.9932e-3, 4.4709e2, 7.9328e2,\right]$,\\
$\left[6, 4.5260e0, 4.1270e-2, 8.9257e-3, 1.7101e2, 4.8403e2\right]$.\\
We noticed as in the previous cases that the most linear values occur for $1<n<4$. In this approach, we can use the $f_a$ and $f_b$ function similarity of solution  for $k$ to determine the appropriate regime for a reasonable  solution. Here, we notice a sudden departure of similar value between $k_a$ and $k_b$  (about 0.4 difference ) at $n=4$ and so we conclude that the probable ``rate constant'' is about the range given by the values spanning $n=2$ and $n=3$. Interestingly, the $\lambda_{\infty}$  values are approximately similar to the ones for method 1 and 2 for polynomial evaluation 2  and 3 for those methods. More study with reliable data needs to be done in order to discern and select  appropriate criteria that can be applied to these non-linear methods.

\subsection{1.4 Evidence of varying kinetic coefficient $k$} Finally, what of direct methods that  do not assume the constancy of $k$ which was the case in the above subsections? Under the linearity assumption $x=\alpha\lambda(t)$, the rate law  has the form $dc/dt=-k(t)(a-x)$ where $k(t)$ is the instantaneous rate constant and this form implies 
 \begin{equation} \label{eq:13}
 k(t)=\frac{d\lambda/dt}{\lambda_{\infty}-\lambda(t)}
\end{equation}
 If $\lambda_\infty$ is known from accurate experiments or from our computed estimates, then $k(t)$ is determined; the variation of $k(t)$ provides crucial information concerning reaction kinetic mechanism and energetics,  from at least one theory  recently developed for elementary reactions \cite{cgj1} and for such theories and developments, it may be anticipated  that nonlinear methods would be used to accurately determine $k(t)$ that would yield the so-called ``reactivity coefficients'' \cite{cgj1} that account for variations in $k$ that would provide fundamental information concerning activation and free energy changes.  
 \begin{figure}[htbp]
 \begin{center}
	\includegraphics[width=7cm]{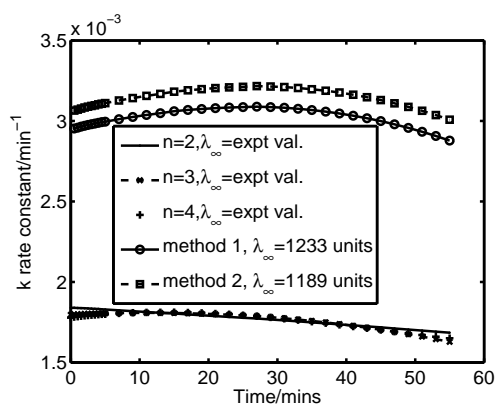}
	\end{center}
		
	\caption{Variation of $k$ with time or concentration changes based on the experimental value $\lambda_{\infty}=2050 \text{units}$ and the computations based on different polynomial degrees $n=2,3,4$ and the computed $\lambda_{\infty}$ values for Method 1 and Method 2 for fixed polynomial degree $n=3$.}
	\label{fig:5}
\end{figure}
 Figure(\ref{fig:5}) refers to the computations under the assumption of first order linearity of concentration and the conductivity. Whilst very preliminary, non-constancy of the rate constants are evident, and one can therefore expect that another area of fruitful experimental and theoretical development can be expected from these results.
\section{2. CONCLUSIONS}
The results presented here provides alternative developments based on NLA that is able to probe into the finer details of kinetic phenomena than what the standard representations  allow for, especially  in the the areas of changes of the rate constant with the reaction environment. Such studies would involve building up another set of axioms that is consistent with a varying $k(t)$ kinetic coefficient. Even with the assumption of invariance of $k$, one can always choose the best type of polynomial order that is consistent with the assumption, and it appears that the initial concentration  as well as the rate constant seems be be predicted as global properties based on the polynomial expansion. 
\section{ACKNOWLEDGMENTS}
This work was supported by Science Faculty conference allocation and grants UMRG(RG077/09AFR),  FRGS(FP037/2008C) and  
 PJP(FP037/2007C) of the Malaysian Government.  \\ \\

\end{document}